\documentclass[twocolumn,showpacs,prb,amsmath,amssymb]{revtex4}

\usepackage{graphicx}% Include figure files
\usepackage{dcolumn}% Align table columns on decimal point
\usepackage{bm}% bold math

\begin{document}

\preprint{APS/123-QED}

\title{Pseudogap behavior of RuP probed by photoemission spectroscopy}% Force line breaks with \\

\author{K.~Sato$^1$}
\author{D.~Ootsuki$^2$}
\author{Y.~Wakisaka$^1$}
\author{N.~L.~Saini$^{3,1}$}
\author{T.~Mizokawa$^{1,2}$}
\author{M.~Arita$^4$}
\author{H.~Anzai$^4$}
\author{H.~Namatame$^4$}
\author{M.~Taniguchi$^{4,5}$}
\author{D.~Hirai$^6$}
\author{H.~Takagi$^{2,6}$}

\affiliation{$^1$Department of Complexity Science and Engineering, University of Tokyo, 5-1-5 Kashiwanoha, Chiba 277-8561, Japan}
\affiliation{$^2$Department of Physics, University of Tokyo, 5-1-5 Kashiwanoha, Chiba 277-8561, Japan}
\affiliation{$^3$Department of Physics, University of Roma "La Sapienza" Piazzale Aldo Moro 2, 00185 Roma, Italy}
\affiliation{$^4$Hiroshima Synchrotron Radiation Center, Hiroshima University, Higashi-hiroshima 739-0046, Japan}
\affiliation{$^5$Graduate School of Science, Hiroshima University, Higashi-hiroshima 739-8526, Japan}
\affiliation{$^6$Department of Advanced Materials, University of Tokyo, 5-1-5 Kashiwanoha, Chiba 277-8561, Japan}

\date{\today}% It is always \today, today,

\begin{abstract}
We have studied the electronic structure of RuP and related Ru pnictides 
using photoemission spectroscopy. Ru 3$d$ core-level and valence-band spectra 
of RuP show that the Ru valence is +3 with $t_{2g}^5$ configuration. 
The photoemisson spectral weight near the Fermi level is moderately 
suppressed in the pseudogap phase of RuP, consistent with the pseudogap 
opening of $2\Delta/k_BT_c$ $\sim$ 3 (gap size $\Delta$ $\sim$ 50 meV 
and transition temperature $T_c$ $\sim$ 330 K). The Ru 3$d$ peak remains sharp 
in the pseudogap phase and the insulating phase of RuP, suggesting that 
the electronic orderings responsible for the phase transitions are 
different from the conventional charge density wave. 
\end{abstract}

\maketitle

Intensive and extensive research activities have been dedicated to understand 
the electronic phase diagram of Fe pnictide family since the discovery of 
superconductivity and magnetism in LaFeAsO$_{1-x}$F$_x$. \cite{1,2} 
The Fe pnictide superconductors show an interesting interplay between 
superconductivity and magnetism which is similar to cuprate superconductors. 
In the case of the cuprate superconductors, pseudogap behavior is believed 
to be one of the key ingredients to understand the relationship between 
the superconductivity and the magnetism. On the other hand, pseudogap behavior 
has never been established in the Fe pnictide superconductors although 
it was claimed in an early photoemission study. \cite{3} Very recently, 
Hirai {\it et al.} have discovered that Ru pnictides have a unique electronic 
phase diagram with insulating, superconducting, and pseudogap phases. \cite{4}
This has opened up a new opportunity to study the relationship between 
superconductivity and pseudogap in the pnictides.
Ru pnictides have a complicated three-dimensional structure (MnP-type structure) 
as schematically shown in Fig. 1 and shows a transition from a metal to 
a non-magnetic insulator. Transitions from metals to non-magnetic 
(or almost non-magnetic) insulators have been reported in various transition-metal 
compounds including pyrochlore-type Tl$_2$Ru$_2$O$_7$, \cite{5,6} 
spinel-type CuIr$_2$S$_4$, \cite{7,8} LiRh$_2$O$_4$, \cite{9} MgTi$_2$O$_4$, \cite{10,11},
hollandite-type K$_2$V$_8$O$_{16}$. \cite{12,13} 
However, superconductivity has not been realized in any of these by destroying 
the non-magnetic insulating phases. In this context, the non-magnetic insulating 
phase of RuP is very interesting and important since the superconductivity 
is induced by Rh doping. \cite{4} 
The electric resistivity of RuP takes a minimum at $T_1$ = 330 K with a drastic 
increase due to the metal-to-insulator transition around $T_2$ = 270 K. 
The magnetic susceptibility of RuP shows Pauli paramagnetic behavior 
above $T_1$ and gradually decreases below $T_1$ with an almost discontinuous 
drop around $T_2$ to a negative value which is comparable to the expected 
core diamagnetism. \cite{4} This suggests that RuP is a normal metal above
$T_1$ and an non-magnetic insulator below $T_2$. In the temperature range 
between $T_1$ and $T_2$, RuP shows the pseudogap behavior. RuAs resembles 
RuP with $T_1$ = 280 K and $T_2$ = 200 K while RuSb is a normal metal 
down to the lowest temperature.
In this paper, we report core-level and valence-band x-ray photoemission 
spectroscopy (XPS) and valence-band ultraviolet photoemission spectroscopy (UPS) 
of RuP and related Ru pnictides. While XPS provides information on the fundamental 
electronic structure of Ru$^{3+}$ ($t_{2g}^5$) in RuP, spectral evidence of 
pseudogap opening is obtained by UPS measurement of RuP.

\begin{figure}
\includegraphics[width=8cm]{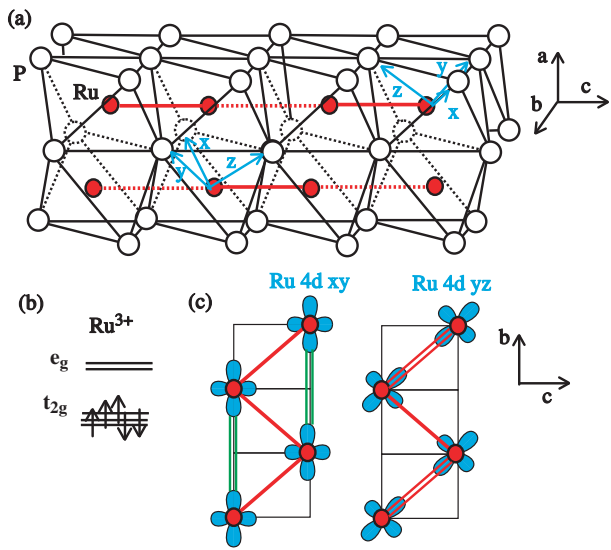}
\caption{(color online) (a) Schematic drawing for crystal structure of RuP. 
The RuP$_6$ octahedra share their edges and faces to form the three-dimensional structure. 
The distortion of RuP$_6$ octahedron is neglected and the shift of Ru ions are included. 
The shorter (longer) Ru-Ru bonds are indicated by the solid (dashed) lines. 
(b) Electronic configurations of Ru$^{3+}$ ($t_{2g}^5$). (c) Ru 4$d$ $t_{2g}$ orbitals 
on the zig-zag double chain formed by the shorter Ru-Ru bonds. In case of Ru 4$d$ $xy$ or $yz$ 
orbital ordering, the Ru-Ru dimers are formed as indicated by the double lines.}
\end{figure}

Polycrystalline samples of RuP, Ru$_{0.75}$Rh$_{0.25}$As, and RuSb were 
prepared as reported in ref. 4. \cite{4} The XPS measurements were performed 
using a JPS-9200 spectrometer equipped with a monochromatized Al K$\alpha$ x-ray source 
(h$\nu$ = 1486.6 eV). The total energy resolution was 600 meV. The base pressure of 
the spectrometer was in the 1.0$\times$10$^{-7}$ Pa range. The UPS measurements were performed 
using SES-100 analyzer with the He I line (h$\nu$ = 21.2eV). The total energy resolution 
was 30 meV. The base pressure of the spectrometer was 5.0$\times$10$^{-8}$ Pa. The high-energy 
resolution UPS measurements have been performed at beamline 9A at Hiroshima Synchrotron 
Radiation Center (HiSOR). The photon energy from the normal incidence beamline 
was set to h$\nu$=10 eV. The total energy resolution was 8 meV. The binding energy 
was calibrated using the Fermi edge of the gold reference sample. 
The base pressure of the chamber was 2.0$\times$10$^{-9}$ Pa. The polycrystalline samples 
of RuP were fractured in situ at 300 K in order to obtain clean surfaces for the XPS and UPS measurements.

\begin{figure}
\includegraphics[width=9cm]{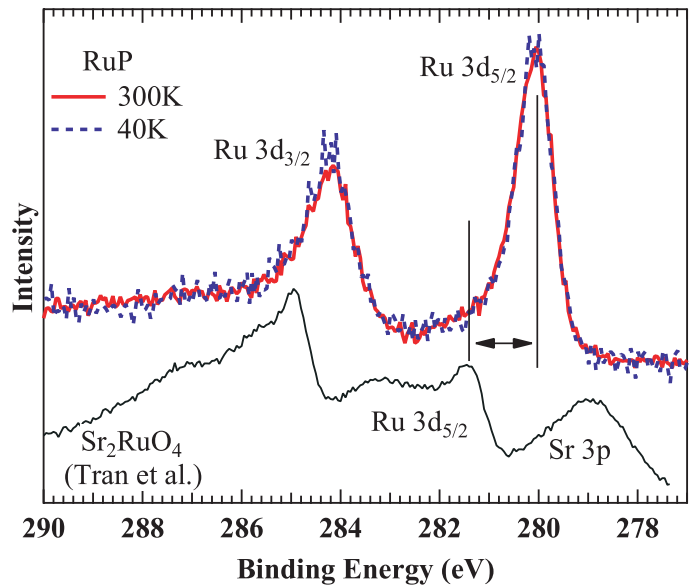}
\caption{(color online) Ru 3d core-level XPS photoemission spectra of RuP taken 
at 300 K and 40 K compared with that of Sr$_2$RuO$_4$ by Tran {\it et al}. \cite{14}}
\end{figure}

Figure 2 shows the Ru 3$d$ core level XPS spectra of RuP measured at 300K and 40 K 
which are compared with the Ru 3$d$ core level XPS spectrum of Sr$_2$Ru$^{4+}$O$_4$ at 300K. \cite{14}
The Ru 3$d_{5/2}$ peaks of RuP and Sr$_2$RuO$_4$ are located at 280 eV and 281 eV, respectively. 
The difference of the Ru 3$d$ binding energy is consistent with difference of Ru valence 
between trivalent RuP and tetravalent Sr$_2$Ru$^{4+}$O$_4$. Therefore, the Ru$^{3+}$ in RuP 
takes $t_{2g}^5$ electronic configuration with S = 1/2. Thus, the mechanism of the metal-to-insulator 
transition of RuP should be different from that of Tl$_2$Ru$_2$O$_7$ with $t_{2g}^4$ electronic 
configuration. \cite{6} The Ru 3$d_{5/2}$ peaks of RuP at 300 K and 40 K are rather narrow and 
do not change appreciably across the metal-to-insulator transition at 270 K. 
If the metal-to-insulator transition is driven by charge density wave (CDW) formation, 
the Ru 3$d$ peak of the CDW phase is expected to be broadened or be split due to the Ru 4$d$ 
charge modulation just like the observations in TaS$_2$ \cite{15} and CuIr$_2$S$_4$. \cite{16}
However, the Ru 3$d_{5/2}$ peak of the insulating phase is narrow and is essentially the same as 
that of the metallic phase, indicating that there is no CDW or that the magnitude of charge modulation, 
if exists, is very small.

Figure 3(a) shows the valence-band XPS (h$\nu$ = 1486.6 eV) and UPS (h$\nu$ = 10 eV) spectra of RuP 
taken at 300 K. The broad structures ranging from the Fermi level ($E_F$) to 2.5 eV are derived from 
the Ru 4$d$ $t_{2g}$ orbitals hybridized with the P 3$p$ orbitals. The near-$E_F$ spectrum below 1.0 eV 
is featureless while the spectral features at 1.4 eV, 1.9 eV, and 2.2 eV are clearly observed. 
Since Ru 4$d$ photo-ionization cross section relative to P 3$p$ increases from 10 eV to 1486.6 eV, 
the spectral weight from $E_F$ to 1.0 eV has stronger Ru 4$d$ character than that from 1 eV to 2.5 eV. 
In Figs. 3(b) and (c), the near-$E_F$ UPS spectrum of RuP taken at 300 K is compared with that 
of Ru$_{0.75}$Rh$_{0.25}$As and RuSb. In Fig. 3(c), the photoemission spectra are divided by 
Fermi distribution functions for each temperature convoluted with a Gaussian function of FWHM of 8 meV. 
The spectral weight from $E_F$ to $\sim$ 50 meV is suppressed in RuP compared to those of 
Ru$_{0.75}$Rh$_{0.25}$As and RuSb. This spectral weight suppression is consistent with 
the pseudogap phase of RuP between $T_1$ = 330 K and $T_2$ = 270 K. The energy scale of 
pseudogap $\Delta$ $\sim$ 50 meV is roughly consistent with the pseudogap transition 
temperature $T_c$ $\sim$ 330 K since the $2\Delta/k_BT_c$ value is $\sim$ 3 close to the BCS value.

Figure 4(a) shows the near-$E_F$ UPS spectra of RuP taken at 300K, 230K 120K and 40K at h$\nu$=10 eV. 
They are normalized to the integrated spectral weight from 0.2 eV to -0.1 eV. 
Across the metal-to-insulator transition temperature, the spectral weight at $E_F$ does not decrease 
in spite of the metal-to-insulator transition. This indicates that the surface region 
of RuP remains metallic even when the bulk undergoes the metal-to-insulator transition. 
Although the main part of the surface-sensitive UPS spectrum does not represent 
the interesting phase transition of the bulk, the electronic structural change of 
the bulk can be picked up in the UPS spectra as discussed below. In order to identify 
the spectral change at $E_F$, the photoemission spectra are divided by Fermi distribution 
functions for each temperature convoluted with a Gaussian function of FWHM of 8 meV. 
In Fig. 4(b), the data up to $3k_BT$ above $E_F$ are plotted where reasonably good
signal-to-noise ratio is available. The spectral weight below $\sim$ 50 meV is suppressed at 300K 
indicating the pseudogap opening of $\sim$ 50 meV. Interestingly, the magnitude of pseudogap 
apparently decreases with cooling. Although the full gap opening associated with 
the metal-to-insulator transition is not observed probably due to the metallic surface,
we speculate that the order parameter of the pseudogap phase is competing with that of 
the insulating phase.

\begin{figure}
\includegraphics[width=9cm]{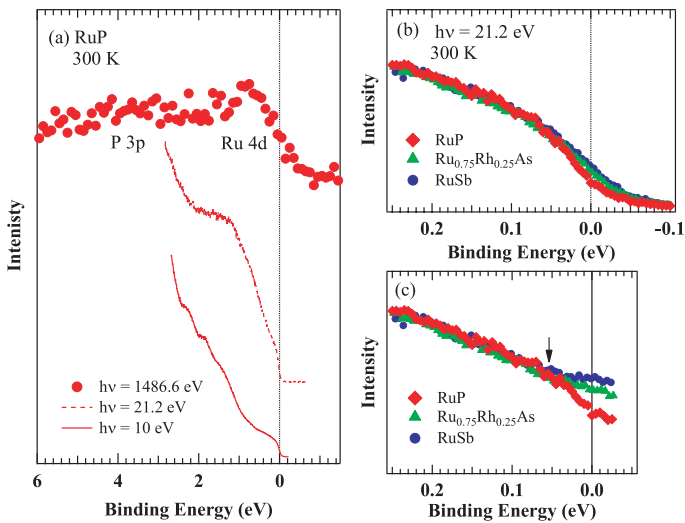}
\caption{(color online) (a) Valence-band XPS and UPS spectra of RuP taken at 300 K. 
(b) Near-$E_F$ photoemission spectrum of RuP compared with those of Ru$_{0.75}$Rh$_{0.25}$As and RuSb. 
(c) Near-$E_F$ photoemission spectra divided by Fermi distribution functions 
for each temperature convoluted with a Gaussian function of FWHM of 8 meV.}
\end{figure}

\begin{figure}
\includegraphics[width=9cm]{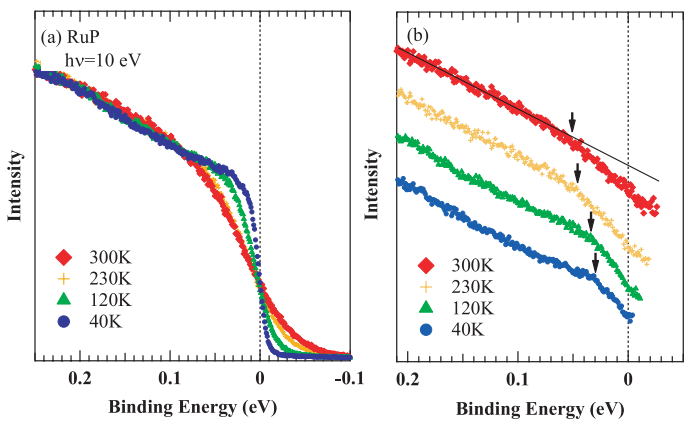}
\caption{(color online) (a) Near-$E_F$ photoemission spectra of RuP taken 
at 300 K, 230 K, 120 K, and 40 K. 
(b) Near-$E_F$ photoemission spectra of RuP divided by Fermi distribution functions 
for each temperature convoluted with a Gaussian function of FWHM of 8 meV.}
\end{figure}

Here, we propose a possible interpretation for the two successive transitions of RuP 
using the idea of the orbitally-induced Peierls mechanism. \cite{17} In the transition from 
the normal metal to the pseudogap phase at $T_1$ = 330 K, since the suppression of the magnetic 
susceptibility is moderate, a kind of band Jahn-Teller effect or an orbital order of the Ru 4$d$ $t_{2g}$ 
bands is expected to produce quasi-one-dimensional Fermi surfaces by the Ru 4$d$ $xy$ orbitals 
[see the left panel of Fig. 1(c)]. In this orbital order, the neighboring Ru 4$d$ $xy$ electrons 
tend to form spin-singlet dimers along the b-axis which can explain the suppression of 
the magnetic susceptibility. However, the Fermi surfaces with the Ru 4$d$ $yz$ and $zx$ 
characters remain below $T_1$. The transition from the pseudogap phase to the insulating phase 
at $T_2$ = 270 K can be assigned to another kind of band Jahn-Teller effect or an orbital order 
to lift the degeneracy of $yz$ and $zx$ orbitals [see the right panel of Fig. 1(c)]. 
The metal-insulator transition at $T_2$ is driven by the Peierls transition of the 
quasi-one-dimensional Fermi surfaces formed by the Ru 4$d$ $yz$ or $zx$ orbital. 
This scenario could be consistent with the fact that the two transitions of RuP 
at $T_1$ = 330 K and $T_2$ = 270 K show different responses to the Rh doping. \cite{4}
While $T_2$ disappears very rapidly in the underdoped region, the pseudogap transition 
temperature $T_1$ remains around room temperature up to the optimum doping level 
where the superconducting transition temperature of Ru$_{1-x}$Rh$_x$P reaches maximum. 

In conclusion, we have studied the electronic structure of RuP using XPS and UPS. 
Ru 3$d$ core-level and valence-band spectra of RuP show that the Ru valence is +3 
with $t_{2g}^5$ configuration. Since the Ru 3$d$ core-level peak remains sharp 
in the pseudogap phase and the insulating phase of RuP, the electronic orderings 
responsible for the phase transitions are different from the conventional charge density wave. 
The photoemisson spectral weight near the Fermi level is moderately suppressed 
in the pseudogap phase of RuP. The energy scale $\Delta$ of the spectral weight sppression is
$\sim$ 50 meV, indicating $2\Delta/k_BT_c$ $\sim$ 3. It is argued that the two successive 
transitions at $T_1$ and $T_2$ in RuP would correspond to the Peierls-like transitions 
in $xy$ and $yz/zx$ orbital channels, respectively.

The authors would like to thank valuable discussions with D. I. Khomskii. 
The synchrotron radiation experiment was performed with the approval of HSRC (Proposal No.11-A-7).

\end{document}